\documentclass[10pt, conference]{IEEEtran}
\IEEEoverridecommandlockouts

\usepackage[numbers, sort&compress]{natbib}
\usepackage{amsmath, amssymb, amsthm, amsfonts}
\usepackage{mathtools}
\usepackage{booktabs}
\usepackage{graphicx}
\usepackage{array}
\usepackage{float}
\usepackage{algorithm}
\usepackage{algpseudocode}
\usepackage{dblfloatfix}
\usepackage{titlesec}
\titlespacing*{\section} {0pt}{1.5ex}{1ex}
\titlespacing*{\subsection} {0pt}{1ex}{1ex}

\AtBeginDocument{%
    \setlength\intextsep{3pt}%
    \setlength\textfloatsep{5pt}}
\begin{document}

\title{Digital Twin Enabled Data-Driven Approach for Traffic Efficiency and Software-Defined Vehicular Network Optimization} 

\author{\IEEEauthorblockN{
Mohammad Sajid Shahriar\IEEEauthorrefmark{1},
Suresh Subramaniam\IEEEauthorrefmark{2},
Motoharu Matsuura\IEEEauthorrefmark{3}, 
Hiroshi Hasegawa\IEEEauthorrefmark{4}, \\ and Shih-Chun Lin\IEEEauthorrefmark{1}, 
}
\IEEEauthorblockA{\IEEEauthorrefmark{1} Intelligent Wireless Networking Lab (iWN), Department of Electrical and Computer Engineering, \\
North Carolina State University, NC, USA. Email: \texttt{\{mshahri, slin23\}@ncsu.edu}\\
\IEEEauthorrefmark{2} George Washington University, WA, USA. Email: \texttt{suresh@gwu.edu}\\
\IEEEauthorrefmark{3} University of Electro-Communications, Tokyo, Japan. Email: \texttt{m.matsuura@uec.ac.jp}\\
\IEEEauthorrefmark{4} Nagoya University, Nagoya, Japan. Email: \texttt{hasegawa@nuee.nagoya-u.ac.jp}}

\thanks{This work was supported in part by the North Carolina Department of Transportation (NCDOT), the National Science Foundation (NSF)
under Grant CNS-221034, and Meta 2022 AI4AI Research.}
}

\maketitle
\begin{abstract}
In the realms of the internet of vehicles (IoV) and intelligent transportation systems (ITS), software defined vehicular networks (SDVN) and edge computing (EC) have emerged as promising technologies for enhancing road traffic efficiency. However, the increasing number of connected autonomous vehicles (CAVs) and EC-based applications presents multi-domain challenges such as inefficient traffic flow due to poor CAV coordination and flow-table overflow in SDVN from increased connectivity and limited ternary content addressable memory (TCAM) capacity. To address these, we focus on a data-driven approach using virtualization technologies like digital twin (DT) to leverage real-time data and simulations. We introduce a DT design and propose two data-driven solutions: a centralized decision support framework to improve traffic efficiency by reducing waiting times at roundabouts and an approach to minimize flow-table overflow and flow re-installation by optimizing flow-entry lifespan in SDVN. Simulation results show the decision support framework reduces average waiting times by 22\% compared to human-driven vehicles, even with a CAV penetration rate of 40\%. Additionally, the proposed optimization of flow-table space usage demonstrates a 50\% reduction in flow-table space requirements, even with 100\% penetration of connected vehicles.

\end{abstract}


\setlength{\abovecaptionskip}{3px}
\setlength{\belowcaptionskip}{-5px}

\section{Introduction}

The integration of CAVs into transportation systems precedes a transformative era, improving comfort, safety, and traffic efficiency. Advancements in autonomous vehicles (AVs) and vehicle-to-everything (V2X) communications offer significant potential for optimizing CAV traffic.

Managing roundabouts with information and communication technology (ICT) is emerging as an effective and safe traffic solution \cite{el2023traffic}. However, the rise of AVs introduces challenges in their efficient navigation, highlighting the need for robust decision-making frameworks. The DT concept has gained attention for its role in data-driven decision-making \cite{NguyenDT}. As an accurate digital replica of its physical counterpart, namely physical twin (PT), the DT operates alongside the PT throughout its lifecycle, optimizing system management and performance, enhancing sustainability. By enabling real-time communication between digital models and physical entities, the DT supports effective engineering decisions and detailed digital representations of physical environments.

Managing traffic at roundabouts with CAVs involves centralized vehicle-to-infrastructure (V2I) and decentralized vehicle-to-vehicle (V2V) systems using road side units (RSUs) and on-board units (OBUs) \cite{islam2023connected}. V2V systems face challenges due to the brief time vehicles spend near the roundabout, requiring precise control logic \cite{campi2023roundabouts}. On contrary, V2I systems need efficient network management strategies, achievable through software-defined networking (SDN) by decoupling control and data planes, enhancing management, programmability, flexibility, and resource optimization. In 5G networks, SDN supports connected ITS and cellular vehicle-to-everything (C-V2X) communications, essential for integrating vehicular networks into the 5G framework \cite{lin2021sdvec}. However, a significant gap exists in SDVN literature, crucial for C-V2X architecture and intelligent transportation systems. This gap is due to the high mobility, dynamic topology, and periodic data transmissions of vehicular networks, requiring unique management strategies and high reliability for safety-critical applications. Thus, research on ensuring reliable network operation to achieve strict Quality of Service (QoS) is essential to address the dynamic nature and environmental factors impacting vehicular networks \cite{lin2021sdvec}.

To enhance CAV traffic efficiency, we propose a data-driven decision support system that leverages edge computing, V2I communications, and digital twin technology. We also introduce a novel method for optimizing SDN flow-table space using data from DT to ensure reliable SDVN operation. The integration of DT with these technologies demonstrates the effectiveness of virtualization and data-driven approaches in transportation and vehicular networks. The paper is structured as follows: Section II reviews related works, Section III details our proposal and its technical components, Section IV presents the simulation settings and experimental results, and Section V offers conclusions and future research directions.

\begin{figure}
  \centering
  \includegraphics[width=1.00\linewidth]{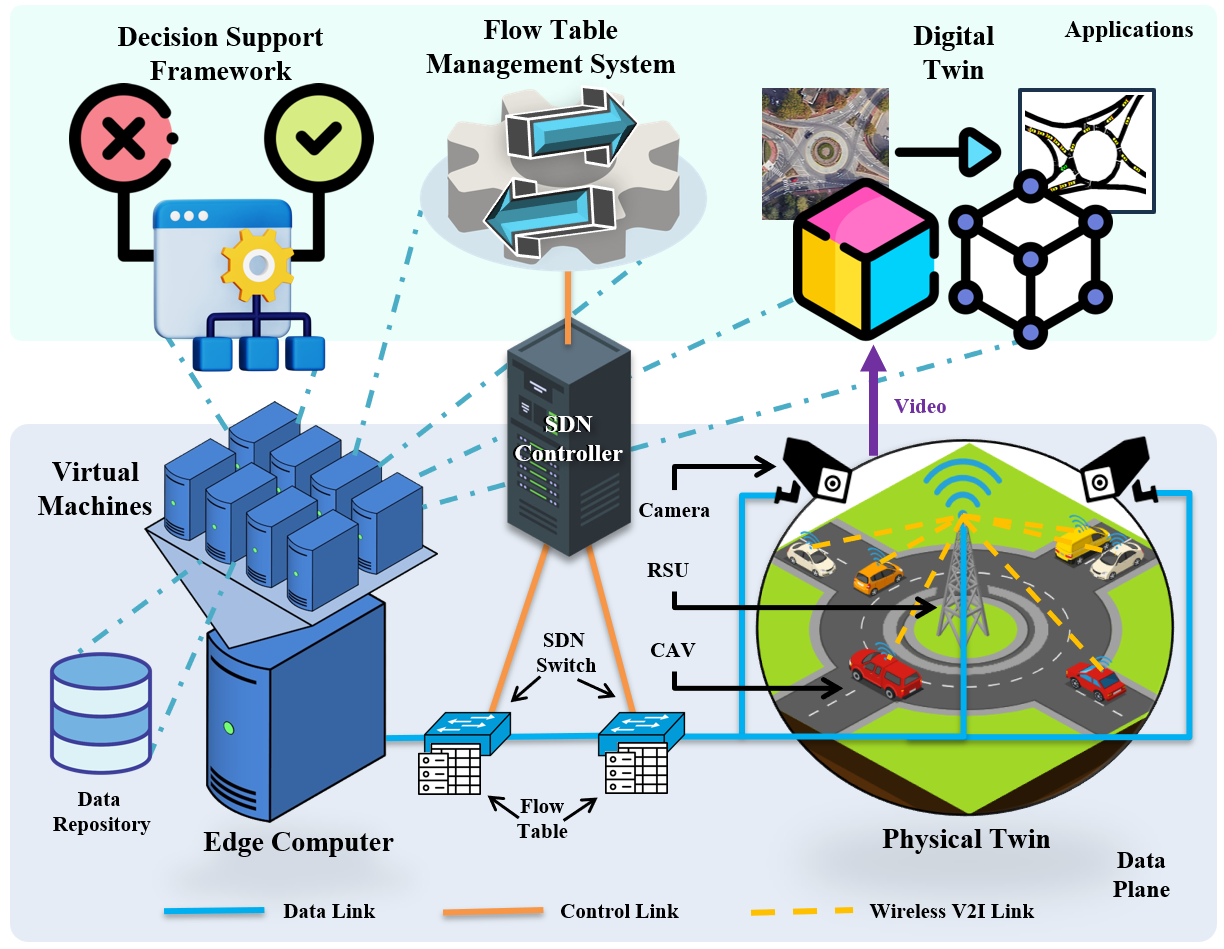} 
  \caption{System model for implementing data-driven edge solutions utilizing digital twin and software defined network.}
  \label{fig:system_model} 
\end{figure}

\section{Related Work}

The integration of CAVs into traffic systems has led to studies focusing on optimizing traffic flow and safety at roundabouts using centralized V2I communication architectures. For example, \cite{el2023traffic} proposes a sustainable system architecture for CAVs at roundabouts, simulating V2I communications where a central signaling unit manages vehicle positions, speeds, and destinations for efficient traffic coordination. Similarly, \cite{martin2021traffic} introduces the Roundabout Manager, a central controller for single-lane roundabouts that prioritizes incoming vehicles and adjusts their trajectories to fit the roundabout's geometry. These studies demonstrate significant advancements in the use of V2I communication and centralized control systems to manage CAV traffic at roundabouts. However, the full potential of data-driven approaches for such systems has yet to be realized. Consequently, these studies do not adequately address the management of the underlying vehicular network needed to support such controllers. Inefficient network management can cause communication failures, scalability issues, inconsistent performance, and increased communication latency, which are major barriers for the implementation of these traffic management systems.

In the field of SDN, various strategies have been devised to enhance the management of flow tables. One notable study in \cite{choi2021udp}, introduces a comprehensive policy for the timely eviction of inactive flow entries, thereby optimizing flow table usage. This approach involves periodically sampling the statistics of User Datagram Protocol (UDP) flow entries to identify and evict inactive entries early. Traffic-based experiments reveal that this system reduces the number of overflow occurrences and the need for flow entries re-installation when compared to random and first-in-first-out (FIFO) policies. In case of hybrid network scenarios, \cite{paliwal2022effective} presents a solution where external boundary forwarding devices of the service provider network are replaced with SDN devices, while internal forwarding devices continue to operate traditionally. The architecture employs a policy-based routing algorithm that efficiently utilizes free IP addresses from the available IP pool, thereby making effective use of the limited flow table space inherent in SDN architecture. Similarly, the research in \cite{shen2020aftm} integrates dynamic timeout assignment with proactive eviction to manage limited flow table resources. The timeout assignment module adjusts timeouts based on flow characteristics, quickly removing short-lived flow entries and retaining entries for flows with large packet intervals. The proactive eviction module works collaboratively to prevent table overflow by removing rule entries when space is constrained. 

Although these studies provide valuable insights into SDN and flow table management, none of them specifically tackle the distinct requirements and challenges posed by vehicular networks. Vehicular networks present distinct characteristics such as high mobility, dynamic and frequently changing topology which necessitate specialized solutions. In this context, our work aims to fill this gap by focusing on flow-table over flow and flow re-installation within vehicular networks. 

\begin{figure}[b]
  \centering
  \includegraphics[width=1.00\linewidth]{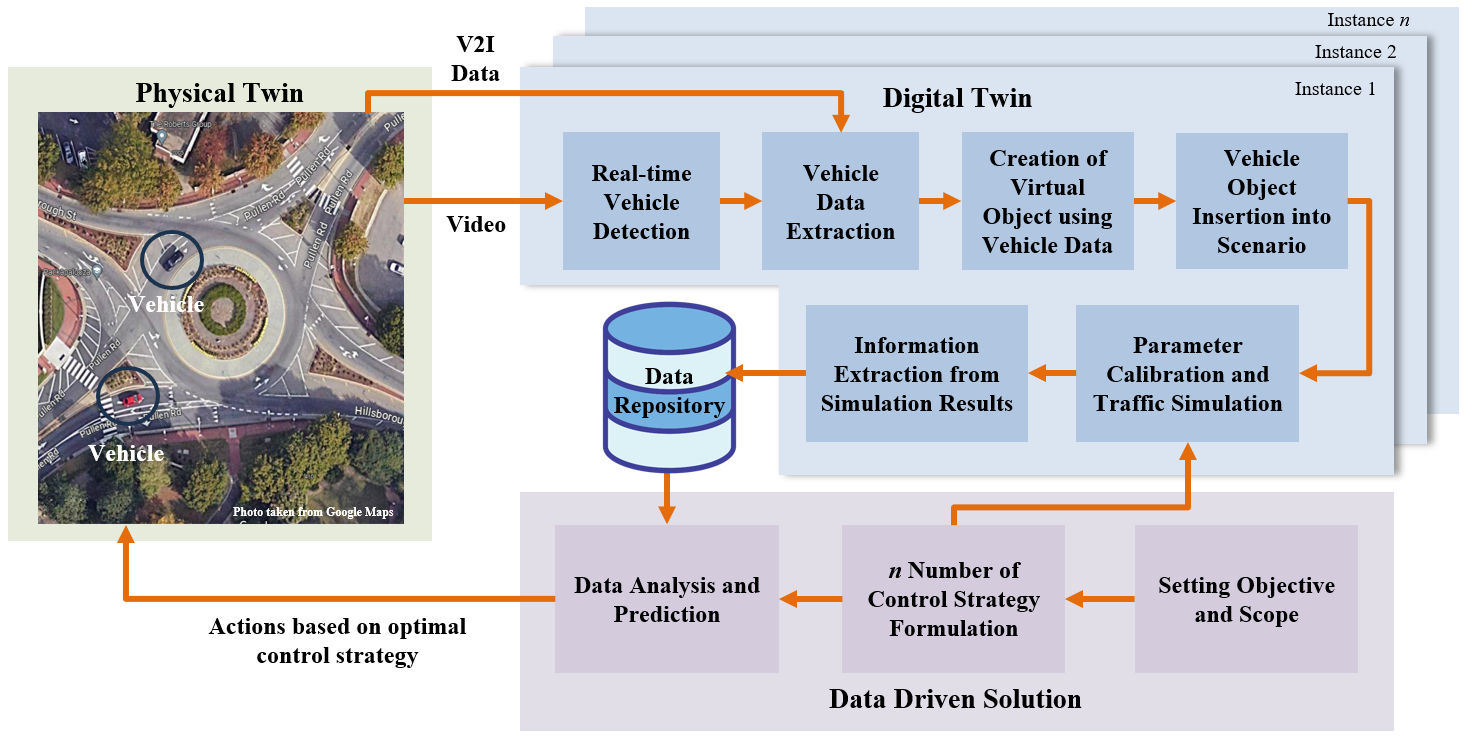} 
  \caption{Proposed digital twin to utilize real-time data and simulation for data driven approaches.}
  \label{fig:digital_twin} 
\end{figure}

\section{Data Driven Solutions Utilizing Digital Twin and SDVN}

To implement a data-driven approach using a DT and SDVN, we design a platform shown in Fig. \ref{fig:system_model}. This platform includes a PT of a roundabout, a DT for virtualization, a decision support framework for CAVs, and a flow-table space optimization system to control SDN switch entries and reduce overflow and re-entries. We propose using SDVN to enable low-latency communication between connected vehicles (CV) and applications on virtual machines (VMs) within the EC, and to manage basic safety message (BSM) data packets via SDN-enabled switches. The SDN architecture consists of a data plane with CAVs, cameras, RSUs, EC, and SDN switches, and a control plane for data flow orchestration. We develop a DT to replicate and simulate the traffic flow of PT and create two data-driven solutions aimed at enhancing traffic efficiency and network reliability. The DT continuously monitors and simulates real-time scenarios using data from the PT, while the decision support framework provides instructions to CAVs via V2I links from RSUs.

\begin{algorithm} [t] 
\caption{Decision provisioning for CAVs to improve traffic efficiency.}
\begin{algorithmic}[1]
\For{$t=1$ \textbf{to} $T$}
    \If{$t \% s == 0$}
        \State Calculate $ y_1 = \arg\max_{x \in H} wt(x) $ 
        \If{$y_1 = N_{in}$ \textbf{or} $y_1 = S_{in}$}
            \State $y_2 = \{ N_{in}, S_{in} \}$
        \Else
            \State $y_2 = \{ E_{in}, W_{in} \}$
        \EndIf
        \State Calculate $ y_3 = H - \arg\min_{x \in H} wt(x) $  
        \State Determine $ y^* = \arg\min_{y \in \{y_2, y_3, H\}} sim_{wt}(y) $ 
        \State Get $CAV = \{v_1, v_2, \ldots, v_i, \ldots, v_n\}$ 
        \For{$i=1$ \textbf{to} $n$}
            \If{ $v_i \in V_{y^*}$}
                \State Send `yield' signal to $v_i$
            \Else
                \State Send `wait' signal to $v_i$
            \EndIf
        \EndFor
    \EndIf
\EndFor
\end{algorithmic}
\end{algorithm}

\subsection{Digital Twin of Roundabout and Connectivity}

Fig. \ref{fig:digital_twin} illustrates the operation of the proposed DT for an abstracted data driven solution. As input of DT, real-time videos of the target PT are used. These videos are captured using cameras connected to the platform. To detect vehicles in the scenario and assess their states in real-time, we employ a video-based vehicle identification and data extraction method facilitated by a proprietary computer vision software.
When a vehicle enters into the video, the software tracks the vehicle and determines the vehicle's category, location i.e. the lane or approach, speed, and timestamp. The DT system connected with this software using an application programming interface (API), uses this information to create a virtual data object (VO) that simulates the vehicle. As such software cannot classify human driven vehicles (HV) and AV yet, we propose using V2I connectivity to identify CAVs and synchronize their data with the DT-generated VOs based on CAV location. The VO is then inserted into traffic and vehicular networking simulation and updated in real-time until the vehicle leaves the video. This method simulates all the vehicles' movements and network data flow in DT. Subsequently, critical data such as travel time, waiting time, average speed, queue length, state of connectivity and networking devices, network statistics etc. are extracted from the simulation results and stored in a data repository. From the data-driven solution perspective, this information is used to compare algorithms, control and optimization strategies in real-time, conduct ``what-if" analyses, and make predictions in both transportation and vehicular networking domains. To implement each strategy or algorithm, a new DT instance is created, allowing key parameters of vehicles, road infrastructure, CAV models, and vehicular networking and connectivity to be modified. All the instances of the DT run simultaneously and as with the previous process, data is extracted and stored in the repository. Once the data-driven solution performs comparative analysis and selects the optimal strategy or algorithm, it can either transform this into actions to be  executed or inform the stakeholders through output message. 
\begin{figure}[b]
  \centering
  \includegraphics[width=1.00\linewidth]{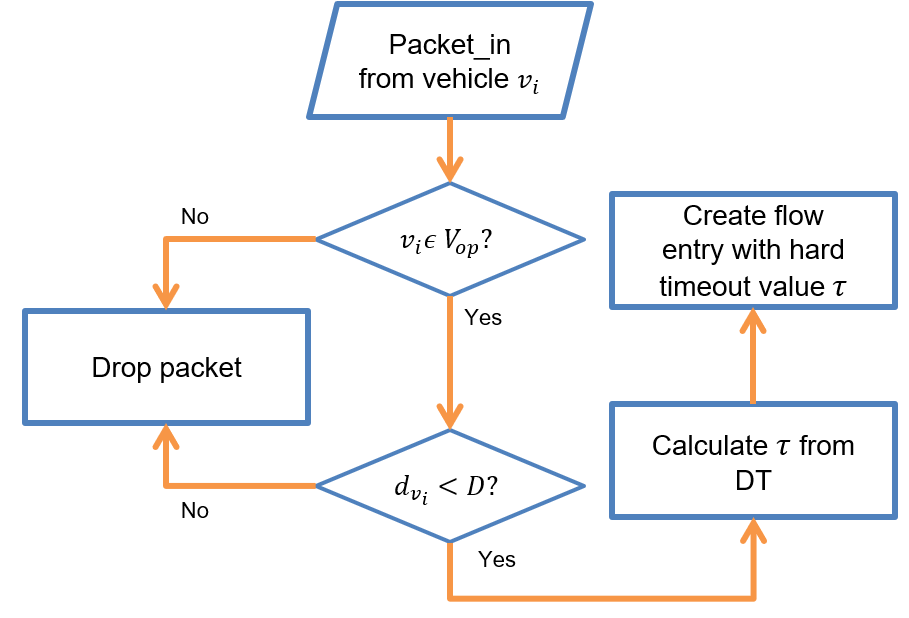} 
  \caption{Flow entry management algorithm for reducing overflow and re-installation in SDVN.}
  \label{fig:algorithm_2} 
\end{figure}

\begin{figure}
  \centering
  \includegraphics[width=1.00\linewidth]{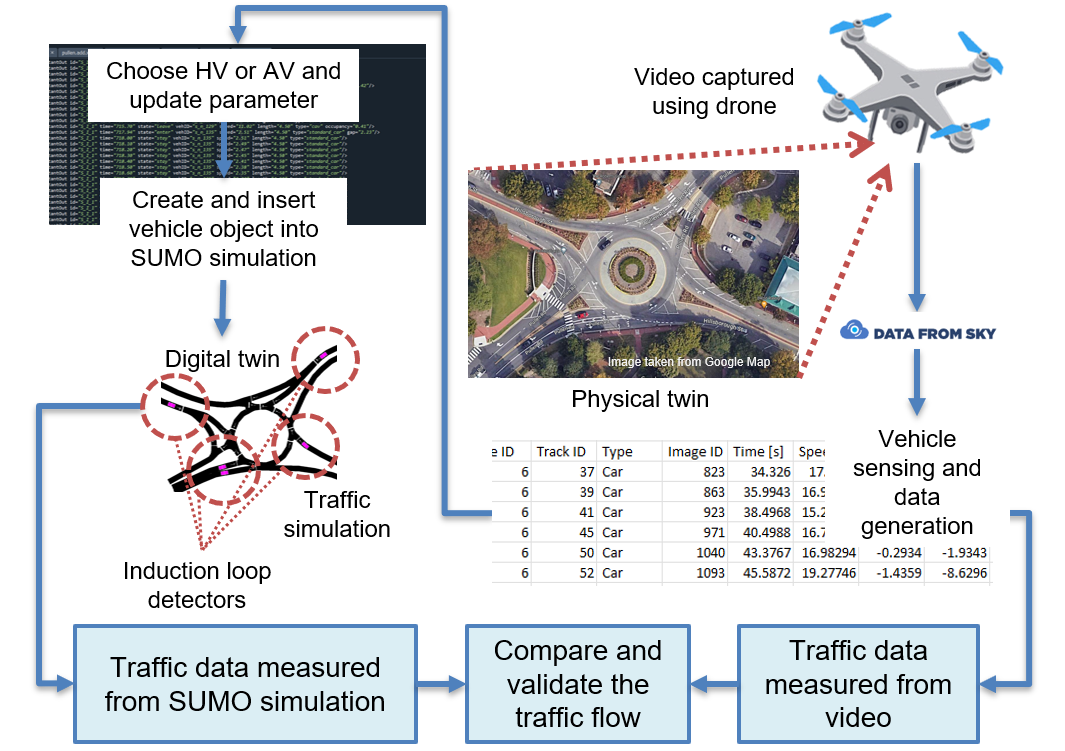} 
  \caption{Creating vehicle-objects in the DT and comparing DT simulation traffic flow with drone-collected data.}
  \label{fig:data_collection} 
\end{figure}

\subsection{Data Driven Decision Support Framework}
The decision-support framework of the platform utilizes the data gathered from the PT to make decisions for the CAVs. This framework considers the four approaches of the PT, where the inbound lanes are defined by a set $H$ where,

\begin{equation}
H = \{ N_{in}, S_{in}, E_{in}, W_{in} \}
\label{eq:set of lanes}
\end{equation}

In equation \ref{eq:set of lanes}, $N_{in}, S_{in}, E_{in}, \text{and } W_{in}$ represent the inbound lanes of North, South, East and West approaches respectively. Assuming the framework is employed for $T$ time period, it gathers data from PT every $s$ time period, to observe the state of PT, such as number of vehicles in the scenario, waiting time of each inbound lanes and so on. The specific objective of this framework is to minimize the average waiting time of vehicles as stated in equation \ref{eq:objective}.

\begin{equation}
\text{minimize } \frac{\sum_{t=1}^{T} \sum_{h \in H} \sum_{i=1}^{n_h} {WT}_{v_i}}{\sum_{t=1}^{T}\sum_{h \in H} n_h}
\label{eq:objective}
\end{equation}

Equation \ref{eq:objective} represents average accumulated waiting time of all the vehicles belong to inbound lanes, where $WT_{v_i}$ is the waiting time of vehicle ${v_i}$ which belongs to lane $h$ and $n_h$ represents the number of vehicles on lane $h$ at time $t$. To achieve the objective, this framework disseminates `yield' or `wait' action messages to the CAVs according to Algorithm 1. The algorithm utilizes a function $wt$ to determine the cumulative waiting time of the vehicles on an inbound lane. The framework quantifies the waiting time of a vehicle using the duration of a vehicle’s stationary state \cite{shahriar2023drl}. To accomplish this, it utilizes speed-data of vehicles acquired from the DT simulation. After calculating the accumulated waiting time for each inbound lane, various strategies involving different lane combinations i.e. $H$, $y_2$ and $y_3$ are simulated in the DT instances to facilitate CAV passage. From these simulations, the average waiting time $sim_{wt}$ for each strategy over the next $p$ time period is extracted. The strategy yielding the minimum $sim_{wt}$ value is labeled as $y^*$, and the AVs belong to  set $V_{y^*}$ which is associated with strategy $y^*$, are instructed to yield in the PT roundabout. To identify $y^*$, this framework prioritizes lanes based on their accumulated waiting time and comes up with 3 strategies utilizing specified combinations of lanes as outlined below:

\begin{itemize}
    \item {\textbf{\textit{All the lanes ($H$)}}}: All the inbound lanes of the four approaches. Which is considered as default when the scenario is not very congested and does not need regulate the AVs which are greedy to yield into the roundabout.
  \item {\textbf{\textit{Two lanes with opposing directions ($y_2$)}}}: The lane which has highest waiting time calculated from PT and the lane which is from opposite direction.
  \item {\textbf{\textit{Lanes excluding the one with least priority ($y_3$)}}}: All the lanes excluding the on that has lowest waiting time calculated from.

\end{itemize}

\subsection{Flow-Table Space Optimization for Vehicular SDN}

Short-lived BSM data packets are managed reactively in SDN to enable bidirectional communication between applications and CVs. An increase in CVs can cause flow-table flooding due to limited TCAM capacity, resulting in delays and reduced throughput \cite{mondal2020flowman}. Premature release of flow entries may lead to frequent re-installations and processing delays. To address this, we optimize flow entry and release policies to minimize overflow and re-installations. We simulate three early-release strategies: removing inactive entries based on flow statistics or idle timeouts \cite{choi2021udp}, using a hard timeout for entry removal, and employing an early-release technique for UDP flows to maintain continuous data exchange between CVs and edge-hosted applications.

\begin{figure}[b]
  \centering
  \includegraphics[width=0.95\linewidth]{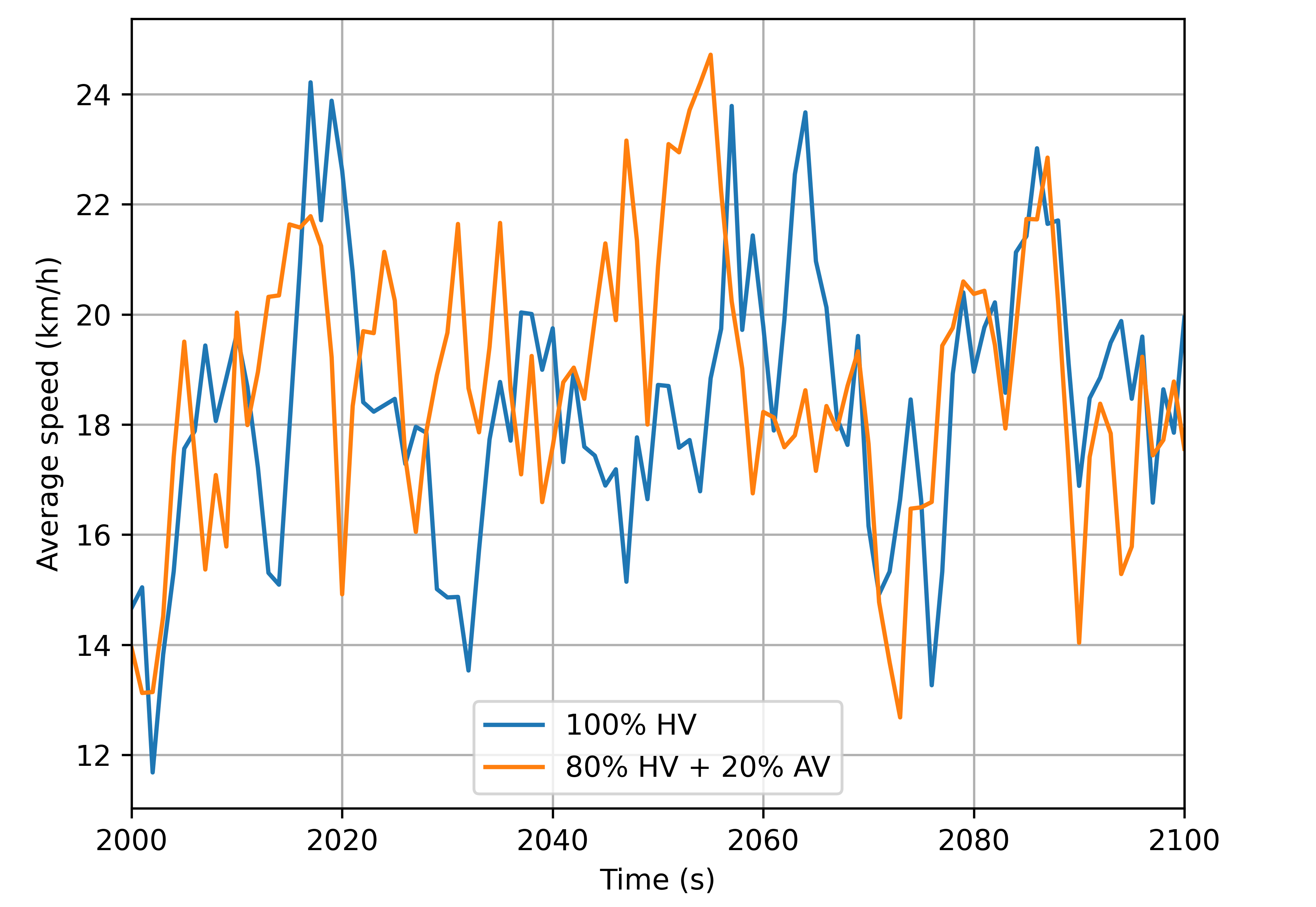} 
  \caption{The discrepancy in average speed observed in the DT simulation underscores the importance of incorporating actual AVs into the PT for precise calibration of AV models.}
  \label{fig:speed_diff} 
\end{figure}

\begin{figure*}
  \centering
  \includegraphics[width=0.95\linewidth]{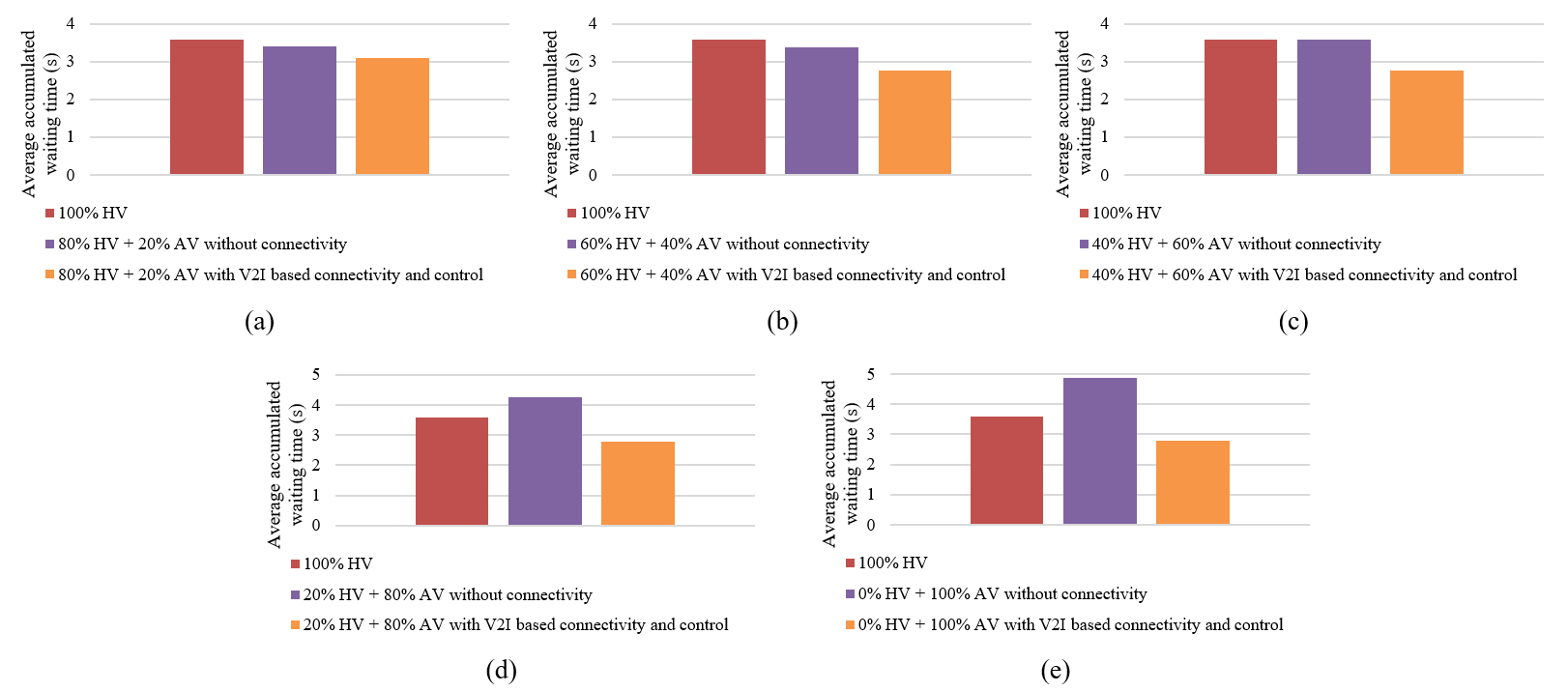} 
  \caption{Effect of increasing number of AVs on average waiting time or traffic efficiency: Sub-figures (a-e) show a 20\% increment in the penetration ratio for each subsequent figure. Initially steady compared to only HV in scenario, efficiency is reduced significantly when AV ratio exceed 50\%, whereas V2I connectivity improves efficiency with higher AV penetration.}
  \label{fig:result_1} 
\end{figure*}

\begin{figure*}[b]
  \centering
  \includegraphics[width=1.00\linewidth]{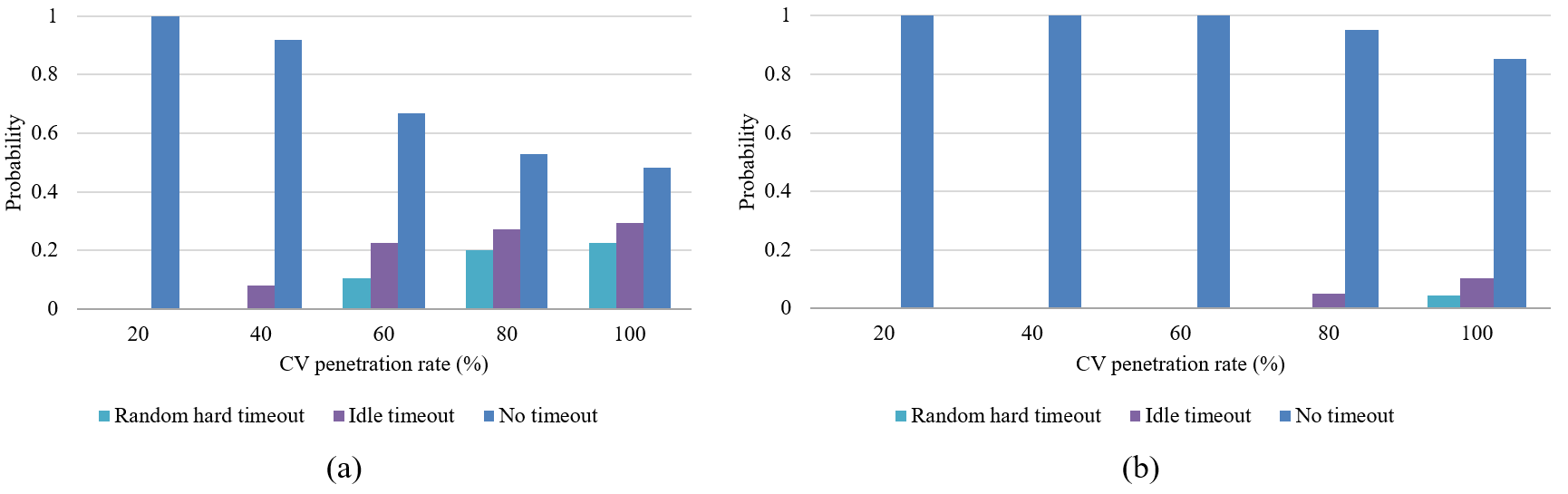} 
  \caption{The overflow probability for managing flow-table entries, with $f_{max}$ set to (a) 125 and (b) 250, demonstrates that reduced flow-table space increases chances of vehicular network failure.}
  \label{fig:result_2} 
\end{figure*}

In vehicular-SDN, a packet\_in event triggers the switch to request a flow setup from the controller if it lacks a rule for an incoming packet from a $v_{CV}$. The controller can either program a rule to direct packets from $v_{CV}$ to the application, use the packet\_out mechanism to route packets through a specific port, or drop them. Properly configuring a rule or flow entry for $v_{CV}$ is crucial to avoid overloading the controller with packet\_out operations, which could disrupt network operations. Since applications need data from $v_{CV}$ to send instructions, each $v_{CV}$ requires 2 flow entries. Each SDN switch has a limited number of flow entries, and different edge applications require separate entries. Assuming a maximum of $f_{max}$ flow entries per application, flow-table overflow occurs if the number of CVs using all applications exceeds half of $f_{max}$. At time $t$, if $V_{CV}$ represents the set of all $v_{CV}$, the number of overflows $N_{of}$ can be determined as shown in equation \ref{eq:overflow}.

\begin{equation}
{N_{of} = 2 |V_{CV}| - f_{max}}
\label{eq:overflow}
\end{equation}

Assuming the set of installed flows $F_t$ in a switch and set of removed flows $R_t$ from the same switch at time $t$, the number of re-installed flows, $N_{re}$ at that time can be calculated using equation \ref{eq:re-install}.

\begin{equation}
N_{re} = | F_t \cap R_t |
\label{eq:re-install}
\end{equation}

\begin{figure*}[t]
  \centering
  \includegraphics[width=1.00\linewidth]{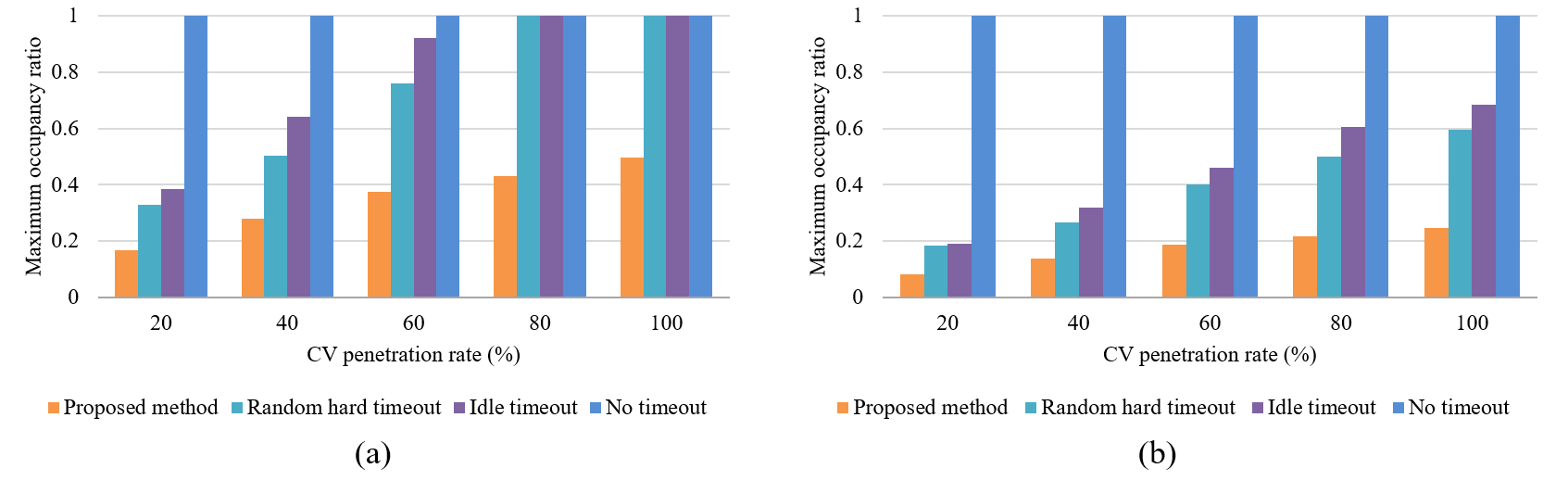} 
  \caption{Maximum occupancy ratio of the flow-table for various strategies, with $f_{max}$ set to (a) 250 and (b) 500, shows optimization of flow-table usage to ensure effective communication between vehicles and applications.}
  \label{fig:result4} 
\end{figure*}

On top of that, as the UDP flows do not ask for connection termination, it is challenging to release the flow entries when they become inactive. So it is necessary to maintain a low table occupancy to ensure overall network operations. To measure the maximum flow entry occupancy for each application, this work implements equation \ref{eq:occupancy ratio}, which determines the maximum value of the ratio between $F_t$ and $f_{max}$ for the time ranging from $1 \text{ to } T$, where $T$ is the current timestamp.

\begin{equation}
R_{oc} = \max_{1 \leq t \leq T} (|F_t|/f_{max})
\label{eq:occupancy ratio}
\end{equation}

This system seeks to accomplish the objective in equation \ref{eq:objective2}, by alleviating $P(of)$, the probability of flow-table overflow and $P(re)$, the probability re-installation of flow entries, while simultaneously reducing $R_{oc}$ and ensuring the existence of essential flow entries to facilitate communication between the application and CVs approaching the inbound lanes.

\begin{equation}
\text{minimize } y = P(of) + P(re) + R_{oc}
\label{eq:objective2}
\end{equation}
\[\text{s.t. } F_t = \{ f_i, f_{i+1} \mid v_i \in V_{op} \}\]
In equation \ref{eq:objective2}, $f_i \text{ and } f_{i+1}$ represent flow entries for CV $v_i$ whereas $V_{op}$ represents set of CVs which are being operated by an application at time $t$. To achieve this objective, the proposed strategy implements a flow entry management algorithm as depicted in Fig. \ref{fig:algorithm_2}. When a packet\_in arrives to the controller from $v_i$, this algorithm first verifies if $v_i$ belongs to set $V_{op}$. If it is, the algorithm proceeds to check if the distance $d_{v_i}$ between $v_i$ and the roundabout is within the operational distance $D$ required by the application. If $d_{v_i}$ exceeds $D$, the algorithm concludes that the flow entry is not necessary and drops the packet. Conversely, if $d_{v_i}$ is less than $D$, the algorithm queries for the maximum travel time $\tau$. It is the required time to traverse the distance $D$ based on the current traffic conditions. $\tau$ is periodically calculated from DT simulation and is stored in data storage which can be retrieved by this system. Following this, two flow entries are established for $v_i$ to facilitate bidirectional communication between an application and $v_i$, each with a hard timeout value set to $\tau$.

\section{Simulation and Result Discussion}
\subsection{Implementation and Experimental Settings}

To develop and evaluate the proposed DT, this study implements the functionalities and conducts unit testing using prerecorded videos of the Pullen Road Roundabout at Pullen Road – Hillsborough Street. As depicted in Fig. \ref{fig:data_collection}, these videos are recorded using an unmanned arial vehicle (UAV) and are fed to the DataFromSky-TrafficSurvey system to simulate real-time video capturing and processing of the DT. Essential information is extracted from the system output to create the vehicle-object, as discussed in the previous section. To replicate the roundabout scenario and conduct traffic simulations, we use OpenStreetMap (OSM) to create a 4-way roundabout road network, excluding pedestrian crossings, based on the Simulation of Urban MObility (SUMO). In the simulation, a variable penetration rate is applied to randomly choose a vehicle and simulate AV with the ACC car-following model. This allows better realization of the proposed data driven applications' scope, the impact of V2I-based connectivity, and to enhance experimental flexibility. To simulate SDN, flow-table, and manage flow-table entries, we use our custom module alongside the Mininet network emulator and Ryu SDN controller. To verify the accuracy of traffic flow simulation by the DT, we measure and analyze the traffic flow from both the PT and DT according to the method in Fig. \ref{fig:data_collection}. In Fig. \ref{fig:speed_diff}, it is evident that due to the AV car-following model, a discrepancy exists between the PT data and the DT simulation data, which can be minimized by optimizing the AV model. For simulation, this study assumes that the total set of vehicles in the scenario consists of HV and AV, where only AVs are connected to the framework through V2I links. Given the penetration ratio of AVs and traffic volume, the number of HVs is calculated using equation \ref{eq:hv_number}, where $Vol_{HV}$ denotes the volume of HVs, $Vol_{total}$ signifies the total number of vehicles per hour in the scenario, and $r_{AV}$ represents the penetration rate of AVs.

\begin{equation}
Vol_{HV} = Vol_{total} - (Vol_{total} * r_{AV})
\label{eq:hv_number}
\end{equation}

To assess the flow management system's efficiency in utilizing flow table space, various $f_{max}$ values are used to simulate flow entry management scenarios, the 3 strategies, idle timeout, random hard timeout, and the proposed strategy are compared 
with no-timeout simulation. Timeout values can be set with flow entries to ensure they are released when the timeout expires. In this study, for random hard timeout, a random value between 10 and 300 seconds is used. In case of proposed method, the operating distance is set to 50 meters. Different levels of AV penetration rates with V2I based connectivity are also incorporated into these experiments to simulate the natural progression of flow entry over time.

\subsection{Result Discussion and Performance Evaluation}

The simulation of the decision support framework demonstrates that increasing the penetration rate of AVs, alongside the traffic flow of HVs, reduces average waiting time. However, once the AV penetration rate exceeds 60\%, the waiting time begins to rise again. Fig. \ref{fig:result_1} shows the relationship between waiting time and penetration rate of AV, aligning with findings in \cite{hajbabaie2022effects}. Nonetheless, the simulation results indicate that average waiting time can be decreased through connectivity between AVs and infrastructure, enabling control from the decision support framework. The EC-based decision support framework, as illustrated, improves traffic efficiency with the increasing penetration rate of AVs, achieving a 22\% reduction in waiting time compared to only HVs, and an additional 18\% reduction in waiting time compared to HVs and AVs, even with a modest 40\% penetration rate of AVs.

\begin{figure}[h]
  \centering
  \includegraphics[width=1.00\linewidth]{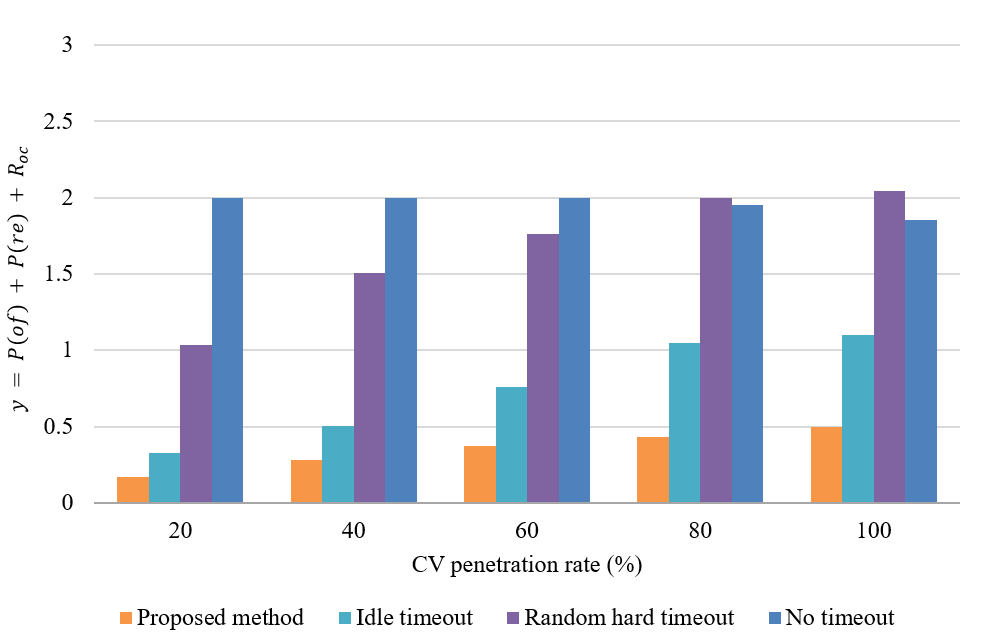} 
  \caption{Reduction of $y$ with $f_{max}$ set to 250 indicates optimized flow-entry lifespan for reliable communication and efficient flow-table utilization.}
  \label{fig:result5} 
\end{figure}

As illustrated in Fig. \ref{fig:result_2}, the probability of overflow increases with the decreasing number of $f_{max}$. However, if no timeout is set for the flows, overflow inevitably occurs, even when $f_{max}$ reaches 250. 
In contrast, when the proposed strategy is implemented, overflow does not occur, even with $f_{max}$ set to 125 and a 100\% penetration rate of CAVs. Flow re-installation is influenced more by the flow entry duration or timeout value than by $f_{max}$, as previously noted. A no-timeout approach prevents flow re-installation since flows are not released. In contrast, a random hard timeout value increases re-installation probability, as CVs continue to transmit packets post-timeout. Implementing an idle timeout minimizes re-installation chances by allowing flow entries to persist longer. The proposed method, with a timeout duration that accommodates vehicle traversal, also shows a low re-installation probability. As shown in Fig. \ref{fig:result4}, $R_{oc}$ decreases with increasing $f_{max}$ in terms of flow table occupancy.
Yet, with no timeout, occupancy is nearly always at its maximum, even with $f_{max}$ set to 500. In this scenario, a random hard timeout performs better than an idle timeout because not all the flows wait for the CVs to stop BSM transmission. Nonetheless, the proposed strategy shows significantly lower occupancy, using only 50\% of the flow table even with a 100\% penetration rate and $f_{max}$ set to 250, which is roughly half the occupancy compared to other strategies. Fig. \ref{fig:result5} illustrates the value of $y$ across various strategies. It indicates that, compared to alternative methods, the proposed strategy decreases chances of table overflow, re-installation, and reduces flow-table occupancy, even under constraints on flow-table size allocation. Moreover, it ensures the creation of flow entries to facilitate effective communication between the applications and CVs.

\section{Conclusions}
Vehicular networks have become a vital technology in ITS, offering solutions to enhance road traffic efficiency. Building on this, our study utilizes SDN and V2I communications to advance improvements in both transportation and vehicular network domains. Specifically, it reduces waiting time at roundabout using proposed data-driven decision support framework aided by digital twin technology to manage AV traffic. Additionally, our proposed SDN approach, designed to accommodate multiple CAV-edge-applications, incorporates a novel flow-entry lifespan optimization method specific to vehicular networks' dynamic attributes. Experimental design and simulation includes calibrating the digital twin simulation with real physical twin data. Results of simulations show that our framework reduces average waiting time by 22\% compared to human-driven vehicles, with a AV penetration rate as low as 40\%. At the same time, our optimized vehicular SDN occupies 50\% less space on flow-tables than other methods, and reduces probability of flow-table overflow and flow re-installation even with full CV penetration. In the future, we plan to conduct integration and system testing on our proposed digital twin to measure the accuracy of real-time AV movement mirroring.

\section*{Acknowledgement}
The authors would like to thank the support from the North Carolina Department of Transportation (NCDOT) under the award number TCE2020-03. The contents do not necessarily reflect the official views or policies of NCDOT.  This paper does not constitute a standard, specification, or regulation.

\begingroup
\footnotesize
\bibliographystyle{IEEEtran}
\bibliography{cite}
\endgroup
\end{document}